\documentclass[12pt]{article}

\usepackage{xr}
\usepackage[thinlines]{easytable}
\usepackage[title]{appendix}
\usepackage[utf8]{inputenc}
\usepackage{amsmath}
\usepackage{amssymb}
\usepackage{physics}
\usepackage{mathtools}
\usepackage{comment}
\usepackage{amsthm}
\usepackage{bbm}
\usepackage{xcolor}
\usepackage{tikz-cd} 
\usepackage{adjustbox}
\usepackage{hyperref}
\usepackage[parfill]{parskip}
\usepackage{soul}
\usepackage[margin=1.3755in]{geometry}
\newtheorem{theorem}{Theorem}[section]

\newtheorem{lemma}{Lemma}[section]

\newtheorem{property}{Property}

\newcommand{\R}{\mathbb{R}}
\newcommand{\set}[1]{\{ #1 \}}
\newcommand{\dprod}[2]{\langle #1, #2\rangle}
\begin{document}
\title{Random Projections of Sparse Adjacency Matrices}
\author{Frank Qiu \thanks{Statistics Department; University of California, Berkeley}}
\date{}

\maketitle
\begin{abstract}
We analyze a random projection method for adjacency matrices, studying its utility in representing sparse graphs. We show that these random projections retain the functionality of their underlying adjacency matrices while having extra properties that make them attractive as dynamic graph representations. In particular, they can represent graphs of different sizes and vertex sets in the same space, allowing for the aggregation and manipulation of graphs in a unified manner. We also provide results on how the size of the projections need to scale in order to preserve accurate graph operations, showing that the size of  the projections can scale linearly with the number of vertices while accurately retaining first-order graph information. We conclude by characterizing our random projection as a distance-preserving map of adjacency matrices analogous to the usual Johnson-Lindenstrauss map.
\end{abstract}

\newpage

\section{Introduction}
The adjacency matrix is a popular and flexible graph representation, encoding a graph's edgeset in an explicit and easy to access manner. Furthermore, many natural graph operations correspond to linear algebraic operations on the adjacency matrix, such as edge addition-deletion, edge or vertex queries, and edge composition. The properties of the adjacency matrix and related quantities like the graph Laplacian also yield important insights about the underlying graph \cite{spectralClustering} \cite{Belkin:2003}.

However, a major defect of the adjacency matrix is that its size scales quadratically with the number of vertices. In many real world applications where the underlying graph is extremely large, this presents a fundamental problem. For example, graphs in the \href{https://snap.stanford.edu/data/}{Stanford Large Network Collection} include graphs derived from social, communication, and road networks that each have millions of vertices. The adjacency matrices of such large graphs would require trillions of parameters to store and work with, which is infeasible in practice. Hence, there has been much interest in graph compression techniques. There is a particular interest in sparse graph techniques, since a majority of real world graphs tend to have sparsely connected vertices. 

In a previous work \cite{qiuTensor}, we introduced a graph embedding method that represents a graph's edgeset as a random sum. This method can be viewed as a random projection technique for adjacency matrices, and in this paper we discuss its application to large sparse graphs. In particular, we show that these random projections have the same space and operational time complexity as popular sparse graph representations like the Compressed Sparse Row format (CSR). Moreover, they retain all of the linear-algebraic graph operations of adjacency matrices, giving a flexible and computationally-efficient representation. Interestingly, the random projection described in this paper is an inner-product preserving projection on the space of adjacency matrices, analogous to the Johnson-Lindenstrauss map for vectors\cite{JLLemma}.

\section{Related Work}
We proposed our random projection method in previous work \cite{qiuTensor}, framing it as a graph embedding method in the spirit of hyperdimensional computing(HDC) \cite{kanerva_sdm}\cite{kleyko_VSA}\cite{Gayler2009ADB}. HDC graph representations encode vertices as high-dimensional vectors, bind these vectors to generate edge embeddings, and sum these edge embeddings to represent the entire edgeset: a bind-and-sum approach \cite{Nickel2016HolographicEO}\cite{Ma2018HolisticRF}\cite{Poduval_graph_embed}\cite{Nunes2022GraphHDEG}\cite{Kang2022RelHDAG}. Our method can be classified as a bind-and-sum method, since we assign a random vector to each vertex and represent an edge by binding the source and target vertices via the tensor(outer) product. However, our method is purely a random projection and does not seek to learn a good node or graph embedding, instead leveraging pseudo-orthogonality to maintain properties of the adjacency matrix.

 Our randomized projections can also be viewed as a "reverse" adjacency matrix factorization. These techniques seek to find an informative and compressed factorization of the adjacency matrix (or adjacency tensor for typed graphs). For example, the RESCAL algorithm \cite{NickelTensor} applied to a $n \times n$ adjacency matrix $A$ learns a $d \times n$ entity matrix $E$ and $d \times d$ relational matrix $R$ such that $A \approx E R E^T$. Here, the columns of $E$ are the entity (vertex) embeddings that encode information about each entity,  and the matrix $R$ encodes information about entity-entity interactions. We work with the source-target factorization of an adjacency matrix $A = C D^T$, where $C$ and $D$ are $n \times k$ matrices whose $i^{th}$ columns are the respective coordinate vectors of the source and target vertex of the $i^{th}$ edge. Our method encodes each vertex as a random vector and then replaces the columns of $C$ and $D$ with their corresponding random vertex codes.

 As a random projection of matrices, our method falls within the field of random numerical linear algebra \cite{martinsson2021randomized}. These are random algorithms and techniques for solving problems in linear algebra, such as solving linear systems and finding maximum/minimum eigenvalues. These techniques aim to offer advantages in speed and memory over their deterministic counterparts, and our method aims to provide both in representing and working with a sparse adjacency matrix. Among these algorithms, the FastRP algorithm\cite{fastRP} is another example of a random projection method. It computes node embeddings by taking a weighted sum of the power of the adjacency matrix and then randomly projects them into a $d \times n$ matrix whose columns form the node embeddings. While the goal of FastRP is to generate node embeddings that capture multi-hop information, our method seeks to informatively embed the adjacency matrix itself.

\section{Intuition Behind the Projection} \label{Sec: Intuition}
Given directed graph $G = (V,E)$ and an arbitrary ordering of its vertex set $V$, its adjacency matrix can be expressed as the sum of outer products of coordinate vectors:
\[
A = \sum_{ij:A_{ij}=1} e_i e_j^T
\]
In this form, it is easy to see that many graph operations correspond to linear operations on the adjacency matrix. For example, adding/deleting the edge $(v_s,v_t)$ corresponds to adding/subtracting the matrix $e_s e_t^T$ from $A$; querying if the graph contains the edge $(v_s,v_t)$ corresponds to the product $e_s^T A e_t$; the $k^{th}$ matrix powers correspond to $k$-length paths. One property these fundamental graph operations share is that they only require orthonormality of the vertex vectors. If we swap the coordinate vectors with any set of orthonormal vectors, the transformed adjacency matrix retains all of its functionality.

For example, let $B = [b_1 \cdots b_n]$ be any orthogonal matrix, whose columns form an orthonormal basis. Changing bases to $B$, our transformed adjacency matrix $A_B$ takes the form:
\[
A_B =  \sum_{ij:A_{ij}=1} b_i b_j^T = B A B^T
\]
In this form, we can perform all the usual adjacency matrix operations by swapping coordinate vectors with their counterparts in $B$. For example, adding/deleting the edge $(v_s,v_t)$ now corresponds to adding/subtracting the matrix $b_s b_t^T$ from $A_B$; the edge query of $(v_s,v_t)$ becomes $b_s A_B^T b_t$; matrix powers still correspond to finite length paths in the sense that  $b_i^T A_B^k b_j = 1$ if and only if there is a $k$-length path from $v_i$ to $v_j$.

From the above discussion, we see that orthonormality is the key property required for exact graph functionality. Hence, relaxing to approximate orthonormality allows us to compress adjacency matrices while retaining graph functionality in a minimally noisy manner.  We make use of random pseudo-orthonormal vectors, which are vectors whose dot products are negligible with high probability. While we can pack at most $d$ orthonormal vectors in a $d$-dimensional space, we can pack many more pseudo-orthonormal vectors in the same space. This allows us to compress adjacency matrices, and in the case of sparse matrices this compression effect is especially pronounced.

\section{Random Projection of Adjacency Matrices}
Consider a directed graph $G = (V, E)$ with $|V|= n$. An ordering of the vertex set $V$ induces the $n \times n$ adjacency matrix $A$:
\[
A_{ij} = \begin{cases}
    1 \quad \text{if $(v_i, v_j) \in E$}\\
    0 \quad \text{otherwise}
\end{cases}
\]
We then construct a random $d \times n$ projection matrix $P_V$, where the columns of $P_V$ are sampled i.i.d. from the uniform distribution on the $d$-dimension unit sphere $\mathbb{S}^{d-1}$. The random projection $\pi_V(A)$ is then given by the following equation:
\begin{equation*}
\pi_V(A) = P_V A P_V^T
\end{equation*}
Expressing the adjacency matrices as the sum of outer products between coordinate vectors $A = \sum e_i e_j^T$, the projection swaps the $i^{th}$ coordinate vector with the $i^{th}$ column $p_i$ of $P_V$: $\sum e_i e_j^T \mapsto \sum p_i p_j^T$. In light of the discussion of Section \ref{Sec: Intuition}, this can be viewed as a pseudo-orthogonal basis change. 

\subsection{Graph Operations on Random Projections}
We saw in Section \ref{Sec: Intuition} that operations in a new orthonormal basis require swapping coordinate vectors with the new basis vectors. Similarly, we can perform all the usual graph operations with the projection $\pi_V(A)$ by substituting the appropriate random code for each vertex. For example, to query if the edge $(v_i,v_j)$ is in the edgeset, we would usually compute the product $e_i^T A e_j$. This returns a 1 if $(v_i,v_j)$ is an edge and 0 otherwise. The randomized analogue then takes the form $p_i^T \pi_V(A) p_j$. This quantity is close to 1 with high probability if $(v_i,v_j)$ is an edge and close to 0 with high probability otherwise. 

In general, every graph operation has an analogue on the projected matrices by substituting in the random vertex codes. Rather than using the coordinate vector $e_i$ associated with vertex $v_i$, when working with the randomized projection $\pi_V(A)$ we use the $i^{th}$ column vector $p_i$ of the projection matrix $P_V$ instead. 

\subsection{Changing the Vertex Set and Graph Aggregation} \label{SubSec:ChangingVertexSet}
One attractive property about our random projections is that they transform nicely under changes to the underlying vertex set. Suppose we wish to expand our graph by adding a new vertex $v_{n+1}$ and new edges $\set{(v_i, v_{n+1})}_i$. With the usual $n \times n$ adjacency matrix, this would require expanding to a $n+1 \times n+1$ matrix and filling in the appropriate entries of the added column and row. However, for the projected matrix we need to only generate a new random vector $p_{n+1}$ and add the appropriate edges to the projection: $\pi_V(A) + \sum_i p_i p_{n+1}^T $. 

Our random matrix projection can be viewed as a projection method for the set of graphs whose vertices are in a fixed vertex set $V$. In light of the above discussion, this projection can be naturally extended to sets containing $V$ or restricted to subsets of $V$ by expanding or restricting the set of random vertex codes respectively. This property is particularly attractive for dynamic graph representations, where the edge and vertex set of the graph change over time. The dimension of the projected matrices stays the same under changes to the vertex set, and the addition/deletion of edges involving new vertices is simple. We only need to keep in mind the total capacity of the projection space: how many edges we can add to a projected matrix before accuracy in graph operations begins to break down. We analyze this behavior in Section \ref{Sec: ScalingProp}.

\subsection{Translation between Different Projections}\label{SubSec:Translation}
Once we assign a random code to each vertex and construct our projection matrix $P_V$, that vertex-vector codebook is fixed. However, we might desire to reassign a new random vector to each vertex.  This translation procedure has a simple analogue for projected matrices. Suppose we have a vertex set $V$ with two different random projection matrices $P_V$ and $Q_V$. The $i^{th}$ columns of $P_V$ and $Q_V$ are the random vectors assigned to vertex $v_i$ under each projection respectively. In order to swap the vector $p_i$ with $q_i$, we construct the translation matrix $T_{(P,Q)} = Q_V P_V^T$. By pseudo-orthonormality, we see that $T_{(P,Q)} p_i \approx q_i$ for every $i$. Hence, if $A_P$ is the projection of $A$ under $P_V$ and $A_Q$ is the projection under $Q_V$, we have the following relation:
\[
A_Q \approx T_{(P,Q)} A T_{(P,Q)}^T
\]

\subsection{Graph Subsets and Aggregation}
Given a subset $S \subseteq V$, the subgraph generated generated by $S$, denoted $G_S \subset G$, is the subgraph generated by all edges involving vertices in $S$. Given the projected adjacency matrix $\pi(A_G)$, we can extract the projection of the subgraph $\pi(A_{G_S})$ by the following procedure. Let $P_S$ be the $d \times |S|$ matrix whose columns are the random codes assigned to each vertex of $S$, and define $T_S = P_S P_S^T$. By pseudo-orthonormality, we have the following approximate relation:
\[
\pi(A_{G_S}) \approx T_S \pi(A_G) T_S^T
\]

This subset procedure, along with graph translation, can be applied to aggregate multiple graphs into a single large graph. For example, suppose we have two disjoint graphs $G$ and $H$. Their aggregate graph $G \cup H = (V_G \cup V_H, E_G \cup E_H)$ is the result of combining their vertex and edge sets together. We saw earlier that edge addition corresponds to adding the projected edges to our projected matrices. Therefore, given projections $\pi(A_G)$ and $\pi(A_H)$, the projection of their aggregate is the sum of their projections: 
\[
\pi(A_G \cup A_H) = \pi(A_G) + \pi(A_H)
\]
This can be combined in tandem with the subsetting procedure to combine selected subgraphs from a collection of graphs. Even when their vertex codes are different, the translation procedure mentioned previously allows use to translate them all into a single code before aggregation.

One interesting application of this suite of procedures is allowing a divide-and-conquer approach to storing a large sparse graph. Graph operations on the projected matrices depend on the size of the projection space (see Section \ref{Sec: ScalingProp}), so breaking a large graph into subgraphs allows us to store their projections in a smaller projection space. Operations involving an individual subgraph also happen in a lower-dimensional space and have decreased time complexity. However, when the need arises to operate on information from a pool of these subgraphs, we can use the above subset and pooling procedures to easily generate a wide suite of projections that correspond to those generated by their graph aggregates.

The above examples demonstrate that we can represent graphs of varying size and different vertex sets in the same projection space. Translation and pooling operations are of fixed dimension, allowing for a unified way to manipulate and combine a broad range of graphs. This suggests that random projections are also ideal for applications where graph aggregation is an important operation such as the divid-and-conquer situation described above.

\section{Scaling Properties of Random Projections} \label{Sec: ScalingProp}
We now study how the size of the random projection space needs to scale with the underlying graph. Importantly, we show that the size scales with size of the edgeset rather than the vertex set. This makes our method particularly amenable to sparse graphs, where the size of the edgeset is proportional to the size of the vertex set.

\subsection{$m$-Order Graph Operations}
First, we need to define $m$-order graph operations. This is important because each order requires a different scaling of the projection space. We say a \textbf{graph operation has order} $\boldsymbol{m}$ if it can be expressed as an operation involving the $m^{th}$ power of the adjacency matrix. For example, the edge query is a first order graph operation because it is a function of the adjacency matrix: $q((v_i,v_j),G) = e_i^T A e_j$.

\subsection{First and Second Order Scaling}
We first characterize how the size of the projection needs to scale in order to retain accurate first and second order graph operations. Intuitively, these correspond to edge information of the 1-hop and 2-hop neighborhoods of the graph respectively. We give two informal statements on how the dimension $d$ of the projection space $\R^{d \times d}$ must scale with the number of edges, subject to a constraint on the vertex connectivity (number of edges each vertex  participates in). An account of the technical results justifying these statements is given in Appendix \ref{AP:ScalProofs}.

\begin{property}[First Order Scaling]\label{Prop:FirstOrderScaling}
       Let $G$ be a graph with $k$ edges with maximum node degree $O(\sqrt{k})$. A random projection into $\R^{d \times d}$ needs to have $d = \Omega(\sqrt{k})$ in order to retain accurate first order operations.
\end{property}

\begin{property}[Second Order Scaling]\label{Prop:SecondOrderScaling}
    Let $G$ be a graph with $k$ edges with maximum node degree $O(k^{1/3})$. A random projection into $\R^{d \times d}$ needs to have $d = \Omega(k^{2/3})$ in order to retain accurate second order operations.    
\end{property}


For sparse graphs, the node degree condition is satisfied for all or a majority of vertices. Indeed, a closer inspection of the proofs in Appendix \ref{AP:ScalProofs} show that as long as the vertices of the relevant edges satisfy this or a much looser bound on the node degree, the results still hold. This allows us to include sparse graphs that might have central vertices, which act like hubs and connect to many other vertices. If we are only interested in first order properties of the graph, the first result implies that we can compress a large sparse adjacency matrix using $n^2$ parameters into a smaller matrix using $\Omega(n)$ parameters, resulting a drastic compression for large sparse graphs.

\subsection{$m$-Order Scaling}
While first and possibly second order operations comprise a bulk of the interesting graph operations, for completeness we have the following informal scaling statement for $m$-order graph operations.

\begin{property}[$m$-Order Scaling]\label{Prop:mOrderScaling}
    Let $G$ be a graph with $k$ edges with maximum node degree $O(k^{1/(m+1)})$. A random projection into $\R^{d \times d}$ needs to have $d = \Omega(k^{\frac{m}{m+1}})$ in order to retain accurate $m$-order operations.
\end{property}


\section{Edge Representations of Sparse Matrices}\label{Sec:Complexity}
In this section, for various sparse matrix representations we analyze both their size complexity as well as the time complexity of graph operations. We also consider the time complexity of numerical sparse matrix methods when appropriate. In this manner, we hope to contextualize both the strengths and weaknesses of our random projections relative to other sparse matrix representations. Table \ref{tab:ComplexityTable} summarizes the results of this section.

\subsection{Alternate Sparse Matrix Representations}
We first consider the coordinate list representation. This represents a sparse matrix as a list of 3-tuples $(r,c,v)$ corresponding to each non-zero entry, where $r$ and $c$ are the row and column indices while $v$ is the value. A related representation is the Dictionary of Keys (DoK) representation, which is similar to the coordinate list representation except now each row-column index $(r,c)$ serves as a key with value $v$. Finally, the CSR format represents a matrix using three arrays containing the non-zero values, their column indices, and the number of non-zero entries above each row respectively. The CSR format can be easily computed from the previous two representations and vice versa.

\subsection{Space Complexity}
A sparse graph with $n$ vertices has $O(n)$ edges. Hence, from their descriptions all three of the alternate sparse matrix representations require $O(n)$ parameters. For random projections, if we are only concerned with representing first-order properties, Theorem \ref{Thm:FirstOrderScaling} establishes that the $d \times d$ projections must have $d = \Omega(n)$, meaning the projections are of size $O(n)$.

\subsection{Graph Operation Complexity}
We analyze some graph operations where the complexity differs based on the representation. For data structure operations, we use their complexity in Python as stated in the \hyperlink{https://wiki.python.org/moin/TimeComplexity}{Python Wiki}. Throughout this section, our random projection are $d \times d$ matrices.

\subsubsection{Edge Query}
To look up if an edge $(v_i,v_j)$ exists in a coordinate list, we merely check if the list contains a tuple whose first two entries are $(i,j)$. Since the coordinate list has length $O(n)$, list lookup has complexity $O(n)$. As a three list format, the CSR representation has the same complexity. However, the "in" operator for dictionaries is $O(1)$, meaning edge lookup in DoK is $O(1)$. The edge query for random projections is the product $p_i^T \pi_V(A) p_j$, which has $O(d^2)$. If our project space is calibrated to preserve only first-order operations, then this edge query has complexity $O(n)$.


\subsubsection{Edge Composition}
Edge composition is a complicated case. For each of the sparse matrix representatiosn, edge compositions requires a nested procedure where, for each edge $(i,j)$, one needs to identify all edges $(j,k)$ and return the edge $(i,k)$, which is naively $O(n^2)$. Importantly, this naive algorithm is not parallelizable, making it especially painful to compute. Alternatively, edge composition naturally corresponds to the second power of the adjacency matrix. For a sparse adjacency matrix with $O(n)$ edges, numerical methods for sparse matrices have complexity at most $O(n^2)$ \cite{SparseCompBorna}.

As for random projections, the naive algorithm for the multiplication of two $d \times d$ matrices has complexity $O(d^3)$\cite{CormenAlgo} \cite{MatComplex}. In order to retain accuracy for a second-order operations, property \ref{Prop:SecondOrderScaling} states that $d = \Omega(n^{2/3})$. Thus, computing the second matrix power of our projections has naive complexity $O(n^2)$ while using methods like Strassen's algorithm\cite{MatComplex} can speed it up even further.

\subsection{Fast Numerical Linear Algebra}
One important aspect of our random projections is they benefit from numerical methods for linear algebra, such as parallel computation. This means that the time complexity of graph operations can be substantially reduced through these methods. The other sparse matrix representations do not enjoy these benefits, as we saw with the edge composition example. Hence, in practice the time complexity of graph operations with random projections can be significantly reduced, and they benefit from advances in numerical linear algebra.

\begin{table}[!htb]
    \centering
    \begin{tabular}{|c|c|c|c|c|c|}
    \hline
     & Random Projections & DoK & CL & CSR & Sparse NLA \\
    \hline
    \hline
    Space Complexity & $O(n)$ & $O(n)$ & $O(n)$ & $O(n)$ & N/A \\[2pt]
    \hline
    Edge Query & $O(n)$ & $O(1)$ & $O(n)$ & $O(n)$ & N/A \\[2pt]
    \hline
    Edge Composition & $O(n^2)^*$ & $O(n^2)$ & $O(n^2)$ & $O(n^2)$ & N/A \\[2pt]
    \hline
    Matrix Addition & $O(n)$ & N/A & N/A & N/A & $O(n)$ \\[2pt]
    \hline
    Matrix Multiplication & $O(n^2)^*$ & N/A & N/A & N/A & $O(n^2)$ \\[2pt]
    \hline
    \end{tabular}
    \caption{\textbf{Table of space/time complexity for various graph and matrix operations.} The underlying sparse graph has $n$ vertices and $O(n)$ edges. From left to right, the methods considered are random projections, dictionary of keys (DoK), coordinate list (CL), compressed sparse row (CSR), and sparse numerical linear algebra (Sparse NLA). For matrix multiplication, we use the time complexity of the naive algorithm rather than other advanced methods \cite{MatComplex}. We assume our random projections are calibrated to first order operations except for edge composition/matrix multiplication($*$) where we assume calibration to second order operations.}
    \label{tab:ComplexityTable}
\end{table}

\newpage

\section{Johnson-Lindenstrauss Analogue for Graphs}\label{Sec:JLGraphs}
    In the introduction, we stated that our random projection method is a norm-preserving map analogous to the one given by Johnson-Lindenstrauss (JL) lemma for vectors. Here, we first discuss a natural inner product on the space of adjacency matrices. We then state results showing how our random projections preserve this inner product and its associated norm for a finite set of adjacency matrices, characterizing our method as a JL map for the space of graphs.

    \subsection{A Natural Inner Product for Adjacency Matrices}\label{Sec:InnerProd}
    In our previous work \cite{qiuTensor}, we noted that there is a natural inner product on the space of adjacency matrices that counts the number of shared edges. Consider two graphs $G$ and $H$ whose vertex sets are contained in some larger set $V$. Given an ordering of $V$, we have the induced adjacency matrices $A_G$ and $A_H$. Define the inner product between their adjacency matrices as:
    \[
    \dprod{A_G}{A_H} \coloneqq tr(A_G^T A_H)
    \]
    Expressing each adjacency matrix as the sum of coordinate outer products, we see that this function counts the number of edges common to both graphs:
    \begin{align*}
    tr(A_G^T A_H) &= tr([\sum_{(i,j) \in E_G} e_i e_j^T]^T [ \sum_{(k,l) \in E_H} e_k e_l^T])\\ 
    &= tr(\sum_{(i,j)\in E_G, (j,l) \in E_H} e_i e_l^T)\\
    &= \sum_{(i,j)\in E_G \cap E_H} 1\\
    &= |E_G \cap E_H|
    \end{align*}
    One can check this is indeed an inner product, and in fact it generates the Frobenius norm: $tr(A^T A) = ||A||_F^2$. 

    Note that we assumed the vertex sets of both graphs were contained in some larger set $V$. This is not an issue for a finite set of graphs $G_i$ with finite vertex sets $V_i$, as we can define $V = \cup V_i$. The adjacency matrix of a graph with respect to this large set $V$ is a block matrix whose single block is equal to the original adjacency matrix. 

    \subsection{Random Projections are a JL Map}
    In light of graph inner product defined in Section \ref{Sec:InnerProd}, we almost have a JL-type result since the Frobenius norm of matrices coincides with the Euclidean norm if we regard matrices in $\R^{d \times d}$ as vectors in $\R^{d^2}$. We confirm this by proving that, with high probability, our random projection method is a map that satisfies the conditions of the JL Lemma.

    \begin{theorem}[Random Projections are a JL map]\label{Thm:JLGraph}
    Consider a set of $N$ graphs with adjacency matrices $A_1,\cdots,A_n$. For small $\epsilon >0$ and $d = \Omega(\frac{\log N}{\epsilon^2})$, with high probability our random projections satisfy:
    \begin{equation}
    (1-\epsilon)||A_i - A_j||^2 \leq ||\pi(A_i)-\pi(A_j)||^2 \leq (1+\epsilon)||A_i - A_j||^2 \label{eq13:JLInequality} \quad \forall i,j
    \end{equation}
    \end{theorem}


\section{Discussion}
We presented a random projection method for sparse adjacency matrices. These random projections exploit pseudo-orthonormality of random vectors to drastically compress the adjacency matrix while still retaining all of its functionality. While the exact scaling depends on the graph properties we wish to preserve, Theorem \ref{Thm:FirstOrderScaling} shows that we can compress $n \times n$ sparse matrices into $\Omega(\sqrt{n}) \times \Omega(\sqrt{n})$ matrices while preserving first order graph operations. These random projections also enjoy properties that their underlying adjacency matrices do not. Sections \ref{SubSec:ChangingVertexSet} and \ref{SubSec:Translation} show that random projection allow us to represent graphs of varying size and different vertex sets in the same projection space. This common space is equipped with modification and aggregation operations that apply to all graphs in the space, and random projections provide a unified way for representing and working with graphs. The complexity analysis of \ref{Sec:Complexity} shows that these random projections are competitive with existing sparse matrix representations and numerical methods, and they can take advantage of numerical techniques for speeding up linear algebra operations. All these properties suggest that random adjacency matrix projections are a dynamic, flexible, and expressive graph compression technique well suited to a variety of applications where large sparse matrices occur. One interesting application of our random compression technique would be to the field of graph neural networks \cite{GNNreview}. Such networks use the adjacency matrix during the linear portion of its computations, and random projections could help extend such networks to both large sparse graphs as well as time-varying, dynamic graphs.


\clearpage

\bibliography{refs.bib}

\begin{thebibliography}{10}

\bibitem{Belkin:2003}
Mikhail Belkin and Partha Niyogi.
\newblock Laplacian eigenmaps for dimensionality reduction and data
  representation.
\newblock {\em Neural Computation}, 15:1373--1396, 2003.

\bibitem{SparseCompBorna}
Keivan Borna and Sohrab Fard.
\newblock A note on the multiplication of sparse matrices.
\newblock {\em Open Computer Science}, 4(1):1--11, 2014.

\bibitem{Boucheron2004}
St{\'e}phane Boucheron, G{\'a}bor Lugosi, and Olivier Bousquet.
\newblock {\em Concentration Inequalities}, pages 208--240.
\newblock Springer Berlin Heidelberg, Berlin, Heidelberg, 2004.

\bibitem{fastRP}
Haochen Chen, Syed~Fahad Sultan, Yingtao Tian, Muhao Chen, and Steven Skiena.
\newblock Fast and accurate network embeddings via very sparse random
  projection, 2019.

\bibitem{CormenAlgo}
Thomas~H. Cormen, Charles~E. Leiserson, Ronald~L. Rivest, and Clifford Stein.
\newblock {\em Introduction to Algorithms, Third Edition}.
\newblock The MIT Press, 3rd edition, 2009.

\bibitem{Gayler2009ADB}
Ross~W. Gayler and Simon~D. Levy.
\newblock A distributed basis for analogical mapping.
\newblock 2009.

\bibitem{JLLemma}
William~B. Johnson and Joram Lindenstrauss.
\newblock Extensions of lipschitz mappings into a hilbert space.
\newblock {\em Contemporary Mathematics}, 26, 1984.

\bibitem{kanerva_sdm}
Pentti Kanerva.
\newblock {\em Sparse Distributed Memory}.
\newblock MIT Press, Cambridge, MA, USA, 1988.

\bibitem{Kang2022RelHDAG}
Jaeyoung Kang, Minxuan Zhou, Abhinav Bhansali, Weihong Xu, Anthony Thomas, and
  Tajana Rosing.
\newblock Relhd: A graph-based learning on fefet with hyperdimensional
  computing.
\newblock {\em 2022 IEEE 40th International Conference on Computer Design
  (ICCD)}, pages 553--560, 2022.

\bibitem{kleyko_VSA}
Denis Kleyko, Dmitri~A. Rachkovskij, Evgeny Osipov, and Abbas~Jawdat Rahim.
\newblock A survey on hyperdimensional computing aka vector symbolic
  architectures, part ii: Applications, cognitive models, and challenges.
\newblock {\em ArXiv}, abs/2112.15424, 2021.

\bibitem{MatComplex}
Yan Li, Sheng-Long Hu, Jie Wang, and Zheng-Hai Huang.
\newblock An introduction to the computational complexity of matrix
  multiplication.
\newblock {\em Journal of the Operations Research Society of China},
  8(1):29--43, 2020.

\bibitem{Ma2018HolisticRF}
Yunpu Ma, Marcel Hildebrandt, Volker Tresp, and Stephan Baier.
\newblock Holistic representations for memorization and inference.
\newblock In {\em UAI}, 2018.

\bibitem{martinsson2021randomized}
Per-Gunnar Martinsson and Joel Tropp.
\newblock Randomized numerical linear algebra: Foundations and algorithms,
  2021.

\bibitem{NickelTensor}
Maximilian Nickel, Xueyan Jiang, and Volker Tresp.
\newblock Reducing the rank in relational factorization models by including
  observable patterns.
\newblock In Z.~Ghahramani, M.~Welling, C.~Cortes, N.~Lawrence, and K.Q.
  Weinberger, editors, {\em Advances in Neural Information Processing Systems},
  volume~27. Curran Associates, Inc., 2014.

\bibitem{Nickel2016HolographicEO}
Maximilian Nickel, Lorenzo Rosasco, and Tomaso~A. Poggio.
\newblock Holographic embeddings of knowledge graphs.
\newblock In {\em AAAI}, 2016.

\bibitem{Nunes2022GraphHDEG}
Igor~O. Nunes, Mike Heddes, Tony Givargis, Alexandru Nicolau, and Alexander~V.
  Veidenbaum.
\newblock Graphhd: Efficient graph classification using hyperdimensional
  computing.
\newblock {\em 2022 Design, Automation \& Test in Europe Conference \&
  Exhibition (DATE)}, pages 1485--1490, 2022.

\bibitem{Poduval_graph_embed}
Prathyush Poduval, Haleh Alimohamadi, Ali Zakeri, Farhad Imani, M.~Hassan
  Najafi, Tony Givargis, and Mohsen Imani.
\newblock Graphd: Graph-based hyperdimensional memorization for brain-like
  cognitive learning.
\newblock {\em Frontiers in Neuroscience}, 16, 2022.

\bibitem{qiuTensor}
Frank Qiu.
\newblock Graph embeddings via tensor products and approximately orthonormal
  codes, 2022.

\bibitem{spectralClustering}
Ulrike von Luxburg.
\newblock A tutorial on spectral clustering, 2007.

\bibitem{GNNreview}
Jie Zhou, Ganqu Cui, Shengding Hu, Zhengyan Zhang, Cheng Yang, Zhiyuan Liu,
  Lifeng Wang, Changcheng Li, and Maosong Sun.
\newblock Graph neural networks: A review of methods and applications.
\newblock {\em AI Open}, 1:57--81, 2020.

\end{thebibliography}
\bibliographystyle{plain}

\clearpage

\begin{appendices}
    \section{Scaling Theorems}\label{AP:ScalProofs}
    We provide theorems justifying the properties listed Section \ref{Sec: ScalingProp}. These are  mainly an adaptation of results found in our previous work \cite{qiuTensor}. Throughout this section, we abuse notation and use the same symbol $u$ to denote both the vertex and its random code. Similarly, we use the same symbol $G$ to denote the graph and its random projection.
    \subsection{Auxiliary Lemma for Spherical Random Vectors}
    We will need the following lemma \cite{qiuTensor}.
    \begin{lemma} \label{Lemma:DPMoments}
        Let $X$ be  the dot product between two vectors sampled uniformly and independently from the $d$-dimensional unit sphere $\mathbb{S}^{d-1}$. Then $E(X) = 0$ and $E(X^2) = \frac{1}{d}$.
    \end{lemma}

    \subsection{Property \ref{Prop:FirstOrderScaling}: First-Order Scaling}
    The following theorem and its proof is an abbreviated adaptation of Theorems 10.7 in our previous work \cite{qiuTensor}. Given an edge $(v_i,v_j)$ and adjacency matrix $A$, the edge query $e_i^T A e_j$ returns 1 if $(v_i,v_j)$ is an edge of the graph and 0 otherwise. Our goal is prove the edge query analogue for our random projection returns the correct value with high probability, showing that our random projection accurately retains first-order edge information. An edge query is a \textbf{true query} if the queried edge exists in the graph and a \textbf{false query} if it doesn't.

    \begin{theorem}\label{Thm:FirstOrderScaling}
        Let $G$ be a graph with $k+1$ edges and maximum node degree $\frac{l}{2}$. For a random projection into $\R^{d \times d}$, we have the following:
        \begin{enumerate}
            \item If $T$ denotes the result of a true query, then:
            \begin{equation}
                P(|T-1| > \epsilon) \leq 2 \exp(\frac{-\epsilon^2}{2( \frac{l}{d} + \frac{k-l}{d^2} + \frac{\epsilon}{3})})\label{eq:EQTrueQuery}
            \end{equation}
            \item If $F$ denotes the result of a false query, then:
            \begin{equation}
                P(|F| > \epsilon) \leq 2 \exp(\frac{-\epsilon^2}{2( \frac{l}{d} + \frac{k+1-l}{d^2} + \frac{\epsilon}{3})})\label{eq:EQFalseQuery}
            \end{equation}
            \end{enumerate}
    \begin{proof}
    Consider the random projection of a graph with $k+1$ edges:
    \begin{equation}
    G = uv^T + \sum_{i=1}^k p_i q_i^T
    \end{equation}
    where $u$, $v$, $p_i$, $q_i$ are random vectors drawn i.i.d. from the uniform distribution on the $d$-dimensional unit sphere.
    
    We first prove equation \ref{eq:EQTrueQuery} by considering the result of querying our random projection for the exist of the edge $(u,v)$, which is a valid edge. We assume $m$ of the other edges share one vertex with $(u,v)$, and WLOG we assume they all share vertex $u$. The result of true query $T$ can be expressed as:
    \begin{equation}
     T = u^T G v = 1 + \sum_{i=1}^m \dprod{p_i}{v} + \sum_{j=1}^{k-m} \dprod{u}{q_j} \dprod{v}{r_j}
    \end{equation}
    From Lemma \ref{Lemma:DPMoments}, both sums have mean 0 with the first sum having variance $\frac{m}{d}$ and the second having variance $\frac{k-m}{d^2}$. Hence, the total variance is $\frac{m}{d} + \frac{k-m}{d^2}$, and applying Bernstein's inequality gives:
    \[
     P(|T - 1| > \epsilon) \leq 2 \exp(\frac{-\epsilon^2}{2( \frac{m}{d} + \frac{k-m}{d^2} + \frac{\epsilon}{3})})    
     \]
     By assumption, we can bound $m \leq \frac{l}{2}$, and plugging in the worst case of $m = \frac{l}{2}$ gives equation \ref{eq:EQTrueQuery}.

    Similarly, now suppose we query by a spurious edge $(s,t)$. Assuming $q$ of the other edges share one vertex with $(s,t)$, we use the same argument as above to derive the false query bound given by equation \ref{eq:EQFalseQuery}.
    \end{proof}
    \end{theorem}
    
    Letting $Q$ denote a general edge query and $\sigma^2 = Var(Q)$, Theorem \ref{Thm:FirstOrderScaling} can be summarized as:
    \[
    P(|Q - EQ| > \epsilon) \leq 2 \exp(\frac{-\epsilon^2}{2(\sigma^2 + \frac{\epsilon}{3})})
    \]
    Rewriting this bound in terms of $\sigma$:
    \[
    P(|Q - EQ| > C \sigma) \leq 2 \exp(\frac{C^2 \sigma^2}{2\sigma^2 + \frac{2\sigma}{3}}) \approx 2 \exp(\frac{-3C^2}{2\sigma})
    \]
    If $l = O(\sqrt{k})$ then $\sigma^2= \frac{O(\sqrt{k})}{d} + \frac{O(k)}{d^2}$. Hence, if $d = \Omega(\sqrt{k})$ then $Q$ is close to $EQ$ with high probability. Since $EQ = 1$ for a true query and $EQ = 0$ for a false query, $Q$ is close to the correct value with high probability. This justifies the statement of property \ref{Prop:FirstOrderScaling}.

    \subsection{Property \ref{Prop:SecondOrderScaling}: Second-Order Scaling}
    This following theorem and its proof is an abbreviated adaptation of Theorems 11.7 from our previous work \cite{qiuTensor}. Two edges are composable if the target vertex of one is the source vertex of another, and an edge $(u,v)$ is in the second power of the adjacency matrix if and only if it is the composition of two composable edges. We need to show that the second matrix power accurately represents edge information. To this end, we prove the edge query returns the correct result with high probability when applied to the second power of the random projection.

    \begin{theorem}\label{Thm:SecondOrderScaling}
        Let $G$ be a graph with composable edges $(u,v)$ and $(v,w)$ along with $k-2$ nuisance edges, and assume $G$ has maximum node degree $\frac{l}{2}$. For a random projection into $\R^{d \times d}$, we have the following results for edge queries involving projection's second matrix power:
        \begin{enumerate}
            \item If $T$ denotes the result of a true query of the second matrix power, then:
            \begin{equation}
                P(|T-1| > \epsilon) \leq 2 \exp(\frac{-\epsilon^2}{2(\frac{2l + 1}{d} + \frac{kl - 3l-3}{d^2} + \frac{k^2 - k(l+2) + l + 1}{d^3} + \frac{\epsilon}{3})}))\label{eq:ECTrueQuery}
            \end{equation}
            \item If $F$ denotes the result of a false query of the second matrix power, then:
            \begin{equation}
                P(|F| > \epsilon) \leq 2 \exp(\frac{-\epsilon^2}{2(\frac{kl}{d^2} + \frac{k^2 - kl}{d^3} + \frac{\epsilon}{3})}))\label{eq:ECFalseQuery}
            \end{equation}
            \end{enumerate}
    \begin{proof}
    For the nuisance edge, assume $m_1$ have common vertex $v$ and $m_2$ have common source $u$ or target $w$, and let $m = m_1 + m_2$. WLOG we may assume all $m_2$ edges have common source vertex $u$, and our projection can be expressed as:
    \begin{equation}
G = uv^T + vw^T + \sum_{i=1}^{m_1} vp_i^T + \sum_{j=1}^{m_2} u p_j^T + \sum_{l=1}^{k-m -2} q_l r_l^T \label{eq4:ECgraph}
    \end{equation}
    From equation \ref{eq4:ECgraph}, we see the second matrix power should only contain the edge $(u,w)$.  
    
    We first prove equation \ref{eq:ECTrueQuery} and query the second matrix power $G^2$ for the presence of the edge $(u,w)$. Letting $\epsilon$ denote the dot product of two independent spherical vectors, the result of this edge query is a sum of terms that are products of i.i.d. $\epsilon$'s. Let $n_1 = k(m_2) - m_2 -m -3$ and $n_2 = k^2 - k(m_2+2)+m_2 +1$. Grouping terms by how many $\epsilon$'s they contain, we write our true query $T$ as:
    \[
    T = 1 + \sum_{i=1}^{m+1} \epsilon_{i}  +   \sum_{j=1}^{n_1} \epsilon_{j_1} \epsilon_{j_2}  + \sum_{h=1}^{n_2} \epsilon_{h_1} \epsilon_{h_2}\epsilon_{h_3}
    \]
    Using Lemma \ref{Lemma:DPMoments}, we see the variance $\sigma^2$ of the error terms is $\sigma^2 = \frac{m+1}{d} + \frac{n_1}{d^2} + \frac{n_3}{d^3}$. An application of Bernstein's inequality gives:
    \[
    P(|T - 1| > \epsilon) \leq 2 \exp(\frac{-\epsilon^2}{2(\frac{m+1}{d} + \frac{n_1}{d^2} + \frac{n_2}{d^3} + \frac{\epsilon}{3})})
    \]
    Since $m_1, m2 \leq l$ and $m \leq 2l$, plugging in the worst case of $m_1 = m_2 = l$ and $m = 2l$ gives equation \ref{eq:ECTrueQuery}.

    Similarly, now we consider querying $G^2$ by a spurious edge $(s,t)$. We assume that $m$ of the edges have either common source vertex $s$ or target vertex $t$. Then, a similar computation as above shows that we can express the false query $F$ as:
    \[
    F = \sum_{i=1}^{km} \epsilon_{i_1} \epsilon_{i_2} + \sum_{j=1}^{k^2 - km} \epsilon_{j_1} \epsilon_{j_2}\epsilon_{j_3}
    \]
    An application of Bernstein's inequality and a worst-case bound gives equation \ref{eq:ECFalseQuery}.
    \end{proof}
    \end{theorem}
    As with Theorem \ref{Thm:FirstOrderScaling}, the bounds of Theorem \ref{Thm:SecondOrderScaling} can be expressed in terms of the query variance $\sigma^2$:
    \[
    P(|Q - EQ| > C \sigma) \leq 2 \exp(\frac{C^2 \sigma^2}{2\sigma^2 + \frac{2\sigma}{3}}) \approx 2 \exp(\frac{-3C^2}{2\sigma})
    \]
    If $l = O(k^{1/3})$, then in both cases all terms in the variance can be expressed as powers of $O(\frac{k^{2/3}}{d})$. Hence, as long as $d = \Omega(k^{2/3})$ then we see that the edge query $Q$ is close to its correct value $EQ$ with high probability. This justifies the statement of Property \ref{Prop:SecondOrderScaling}.

    \subsection{Property \ref{Prop:mOrderScaling}: $m$-Order Scaling}
    We give a proof sketch is adapted from Section 12.1 of our previous work\cite{qiuTensor}. We aim to analyze the accuracy recovering edge information from the $m^{th}$ matrix power of our random projections and how the dimension of the projection space $\R^{d \times d}$ needs to scale with $m$.

    As in the proof of Theorem \ref{Thm:SecondOrderScaling}, let $\epsilon$ denote the dot product of distinct spherical vectors. From the proofs of Theorems \ref{Thm:FirstOrderScaling} and \ref{Thm:SecondOrderScaling}, we need to control the edge query variance to ensure that true and false queries are close to their expected values with high probability.  Intuitively, if the node degree is small then performing an edge query on the $m^{th}$ matrix power will result in a majority of the $k^m$ error terms being products of $m+1$ independent $\epsilon$'s. Such terms will be mean 0 and variance $\frac{1}{d^{m+1}}$. Since the true and false query scores are 1 and 0 respectively, for accurate recovery we need the edge query variance to be less than 1. As a sum of independent terms, the variance of the noise term will be approximately $\frac{k^m}{d^{m+1}}$, implying that $d = \Omega(k^{\frac{m}{m+1}})$ in order to retain accuracy. The node degree bound of $l  = O(k^{1/(m+1)})$ ensures that the variance can be expressed as sums of powers of $O(\frac{k^{m/(m+1)}}{d})$.

    \section{JL Lemma for Graphs}\label{AP:JL}
    Here, we aim to prove Theorem \ref{Thm:JLGraph} by establishing intermediate results. Importantly, along the way we show how our random projection method preserves the graph inner product of Section \ref{Sec:InnerProd}.
    
    \subsection{Random Projections Preserve Inner Products}
    \begin{theorem}\label{Thm:InnerProdPreserve}
    Consider two graphs $G$ and $H$ with $n_1$ and $n_2$ edges respectively. Suppose they have $k$ edges in common. Among all $n_1n_2$ pairs $(e,e') \in E_G \times E_H$, suppose $q$ of these edge pairs share exactly one vertex. Let $\pi(G)$ and $\pi(H)$ denote the random projections of their adjacency matrices $A_G$ and $A_H$ into $\R^{d \times d}$. For any $\epsilon>0$, we have:
    \begin{equation}
     P(|\dprod{\pi(G)}{\pi(H)} - \dprod{A_G}{A_H}| > \epsilon) \leq 2\exp(\frac{-\epsilon^2}{\frac{q}{d} + \frac{n_1 n_2 - k - q}{d^2} + \frac{\epsilon}{3}}) \label{eq9:innerprodBound}   
    \end{equation}
    \begin{proof}
        We can express their random projections as:
        \begin{gather*}
        \pi(G) = \sum^k_{i=1} a_i b^T + \sum_{j=1}^{n_1-k} c_j d_j^T\\ 
        \pi(H) = \sum^k_{i=1} a_i b^T + \sum_{l=1}^{n_2 - k} e_l f_l^T
        \end{gather*}
        where $a,b,c,d,e,f$ are random spherical vectors. Computing their inner product:
        \begin{equation}
        tr(\pi(G)^T \pi(H)) = k + tr(\pi(G)^T \sum_{l=1}^{n_2 - k} e_l f_l^T) + tr(\pi(H)^T \sum_{j=1}^{n_1-k} c_j d_j^T) = k + E \label{eq8:innerprod}
        \end{equation}
        We proceed to bound the error term $E$ in equation \ref{eq8:innerprod}. By assumption, $q$ of the terms in $E$ share one vertex with the remaining terms sharing no vertices:
        \begin{align*}
             E = \sum_{i=1}^{q} \epsilon_i + \sum_{j=1}^{n_1 n_2 - k - q} \epsilon_j \epsilon'_j
        \end{align*}
        where each $\epsilon$ is the dot product of independent spherical vectors. Using Lemma \ref{Lemma:DPMoments} and Bernstein's inequality\cite{Boucheron2004}, we get the following bound:
        \begin{equation*}
        P(|E| \geq \epsilon) \leq 2\exp(\frac{-\epsilon^2}{\frac{q}{d} + \frac{n_1 n_2 - k - q}{d^2} + \frac{\epsilon}{3}})
        \end{equation*}
        \end{proof}
    \end{theorem}
    
    \subsection{Random Projections are a JL Map for Graphs}
    \begin{theorem}\label{Thm:SelfNormPreserve}
        Let $G$ be a graph with $k$ edges. For $d < k $ and small $\epsilon > 0$, its random projection into $\R^{d \times d}$ satisfies:
        \begin{equation}
            P(|\lVert \pi(G)\lVert^2 - k| > k \epsilon) \leq 2 \exp{-d \epsilon^2} \label{eq10:SelfNormBOund}
        \end{equation}
        \begin{proof}
            Of the $k^2$ pairs $(e,e') \in E_G \times E_G$, suppose that $q$ of them share exactly one vertex. Theorem \ref{Thm:InnerProdPreserve} states that:
            \begin{equation*}
                P(|\lVert \pi(G)\lVert^2 - k| > k \epsilon) \leq 2\exp(\frac{-k^2\epsilon^2}{\frac{q}{d} + \frac{k^2 - k - q}{d^2} + \frac{k\epsilon}{3}})
            \end{equation*}
            As $q \leq k^2 -k$, the worst case bound occurs when $q = k^2 - k$ which gives equation \ref{eq10:SelfNormBOund}. Note that if $\epsilon$ is large, the third term $\frac{k\epsilon}{3}$ dominates and gives the trivial bound $2 \exp (-k \epsilon)$.
        \end{proof}
    \end{theorem}
    \begin{theorem}\label{Thm:NormPreserve}
     Let $G$ and $H$ be two graphs with $\lVert A_G - A_H \lVert^2 = m$. For $d < k$ and small $\epsilon > 0$, the random projections of their adjacency matrices into $\R^{d \times d}$ satisfies the following:
     \begin{equation}
        P(| \lVert \pi(G) -\pi(H)\lVert - m| > m \epsilon) \leq  2 \exp(-d \epsilon^2) \label{eq11:sparseNormbound}
     \end{equation}
     \begin{proof}
     If we include signed edges, the graph $G-H$ has $m$ total edges. An application of Theorem \ref{Thm:SelfNormPreserve} gives the result.
     \end{proof}
    \end{theorem}
    \subsubsection{Proof of Theorem \ref{Thm:JLGraph}}
    \begin{proof}
    From Theorem \ref{Thm:NormPreserve}, our random projection satisfies the following inequalities with high probability:
    \begin{equation}
    (1-\epsilon)||A_G - A_H||^2 \leq ||\pi(G)-\pi(H)||^2 \leq (1+\epsilon)||A_G - A_H||^2 \label{eq12:JLBounds}
    \end{equation}
    If we want this to hold for a set of $N$ sparse adjacency matrices, a union bound over all $\binom{N}{2}$ pairs shows equation \ref{eq12:JLBounds} holds with probability at least $1 - 2\binom{N}{2} \exp{-d \epsilon^2} = 1 - N(N-1)\exp{-d \epsilon^2}$. For a fixed probability threshold $T$ of violating the inequalities \ref{eq12:JLBounds}, let us choose the optimal $d$ given $N$, denoted $d_{opt}$. That is, $d_{opt}$ is the smallest integer $d$ such that $N(N-1)\exp{-d\epsilon^2} \leq T$. The optimal $d_{opt}$ satisfies the following inequalities:
    \begin{gather*}
        N(N-1)\exp{-d_{opt}\epsilon^2} \leq T\\
        N(N-1)\exp{-(d_{opt}-1)\epsilon^2} > T
    \end{gather*}
    Combining the two inequalities gives:
    \[
    \frac{\log T - 2 \log N}{\epsilon^2} \leq d_{opt} < \frac{\log T - 2 \log N}{\epsilon^2} + 1
    \]
    Thus, we have $d_{opt} = \Omega(\frac{\log N}{\epsilon^2})$, which matches the scaling of the usual Johnson-Lindenstrauss lemma and establishes our random projections as a JL map for graphs.
    \end{proof}
    \end{appendices}

\end{document}